# Operationalising AI Regulatory Sandboxes under the EU AI Act: The Triple Challenge of Capacity, Coordination and Attractiveness to Providers[*]

**Abstract**

The EU AI Act adopted in 2024 provides a rulebook for all AI systems being put on the market or into service in the European Union. This article investigates the AI Act's requirement that Member States establish national AI regulatory sandboxes for testing and validation of innovative AI systems under regulatory supervision to assist with fostering innovation and complying with the Act's regulatory requirements. Regulatory sandboxes are well-diffused around the globe for fintech, but the AI Act presents a new and unique setting given the broad swathe of risks it covers from health and safety to ethics and fundamental rights. Against the backdrop of the EU's objective that AI regulatory sandboxes would both foster innovation and assist with compliance, considerable challenges are identified for Member States around capacity-building and design of regulatory sandboxes. While Member States are early movers in laying the ground for national AI regulatory sandboxes, the article contends that there is a risk that differing approaches being taken by individual national sandboxes could jeopardise a uniform interpretation of the AI Act and its application in practice. This could motivate innovators to play sandbox arbitrage. The article therefore argues that the European Commission and the AI Board need to act decisively in developing rules and guidance to ensure a cohesive, coordinated approach in national AI regulatory sandboxes. With sandbox participation being voluntary, the possibility that AI regulatory sandboxes may prove unattractive to innovators on their compliance journey is also explored. Confidentiality concerns, the inability to relax legal rules during the sandbox, and the inability of sandbox's to deliver a presumption of conformity with the AI Act are identified as pertinent concerns for innovators contemplating applying to AI regulatory sandboxes as compared with other direct compliance routes provided to them through application of harmonised standards and conformity assessment procedures.

[*] Deirdre Ahern, Professor in Law, Director of the Technologies, Law & Society Research Group, School of Law, Trinity College Dublin. Email: dahern@tcd.ie. The author is grateful to the Fellows of the Milan Information Society Law Centre and the anonymous reviewers for their helpful feedback on this work. The author is a member of Ireland's AI Advisory Council. All views expressed here are made in a personal capacity.

## 1. Introduction

In the EU AI Act[1] (the "AI Act"), the AI regulatory sandbox is conceived as decentralised and polycentric, supporting the testing and compliance needs of intending providers of high-risk AI systems on the EU market. AI regulatory sandboxes are intended to allow short term testing under regulatory supervision to boost competition by aiding regulatory compliant market entry. However, the design and operationalisation of AI regulatory sandboxes by Member States is far from simple. Meanwhile, the question of how they may be received by the market is multi-faceted. The purpose here is thus to provide an overview of salient aspects from the perspectives of Member States and AI innovators and to highlight potential controversies and problems. The article is not arguing for a centralised sandbox environment but rather is advocating that in a decentralised architecture EU leadership around coordination is needed, not least to avoid fragmentation caused by inconsistent messaging to AI innovators on core compliance questions.

First, it is worth acknowledging that constructing EU regulatory and enforcement networks is undoubtedly a complex task (Söderlund & Larsson, 2024). In the case of high-risk systems under the AI Act, decentralised enforcement for high-risk systems by national market surveillance authorities is complemented by centralised coordination through the European AI Office and the supranational AI Board (which has representatives from each Member States), the Scientific Panel and the Advisory Forum. National AI regulatory sandboxes are expected to support the decentralised enforcement model by supporting pre-market compliance preparations of high-risk AI systems.

[1] Artificial Intelligence Act Regulation (EU) 2024/1689 of the European Parliament and of the Council of 13 June 2024 OJ L. 12.7.2024 ("the AI Act").

Regulatory sandboxes are a known phenomenon for the last decade, increasingly employed by fintech regulators the world over to assist innovators to navigate ill-fitting financial services frameworks. However, by contrast, with the AI Act, remarkably the EU adapted the regulatory sandbox and bolted it onto a tailor-made, bespoke legal framework which provides a risk-based legal framework for AI systems being put on the market or into service within the single market.

The AI Act's vision of regulatory sandboxes represents a novel proposition - a network of Member State, regulator-led conduits to assist innovators to navigate regulatory requirements and to expedite their route to market. For Member States, being an early mover on rolling out AI regulatory sandboxes is in part calculated to help make (i) the Member State attractive as a launchpad for AI businesses and (ii) its national AI businesses more competitive, accelerating their path to market by getting ahead in grappling with regulatory compliance with the AI Act. However, to have economic impact, both within Member States and in the Union as a whole, it is essential that AI regulatory sandboxes are perceived as useful and efficient at meeting market needs for providers of high-risk systems. If an AI regulatory sandbox journey is not perceived as useful and efficient, innovators may prefer alternative options to assist their path to market. Accordingly, the objective of this article is to investigate the efficacy of the regulatory design choices in the AI Act in providing for the establishment of AI regulatory sandboxes as a modified version of the fintech regulatory sandbox. The analysis undertaken in this article takes places against the backdrop of the understanding that regulatory design choices matter and that the structural and operational aspects of a system of legal rules constitutes “an intentional, normative activity, whose effectiveness may be measured against their predetermined objectives (Söderlund & Larsson, 2024). As regards the cues in the AI Act, one of the most controversial aspects concerns the extent to which a sandbox is allowed to relax existing rules for participants.

With a view to contributing to scholarly understanding of the evolving operationalisation AI regulatory sandboxes under the AI Act’s architecture, the article advances the state-of-the-art by investigating the following research questions (1) how could regulatory design choices by the EU around decentralised AI regulatory sandboxes affect likely outcomes in practice?; (2) how are Member States responding to the need to design and operationalise AI regulatory sandboxes? (3) Having regard to regulatory design choices in the AI Act what factors could influence AI innovator perception of value or attractiveness in AI regulatory sandbox participation as compared with alternatives? The analysis which follows is informed by and builds its narrative from (a) doctrinal analysis of the AI Act and surrounding EU and national policy documents; (b) the global literature on fintech regulatory sandboxes, (b) data gathered by the author from the public domain concerning the status quo of the 27 Member States’ individual preparations and design choices around AI Act regulatory sandbox operationalisation as of June 2025.[2]

Section 2 provides background, situating the AI Act within its unique regulatory context which bolts a product safety mindset with a fundamental rights lens, and briefly outlining the compliance expectations of providers of high-risks systems. From there, Section 3 is concerned with the conceptualisation of AI regulatory sandboxes. It begins by contextualising the rise of the regulatory sandbox instrument within the fintech regulatory ecosystem, providing a base understanding of the nature of the instrument within regulatory ecology and the lessons learned which inform our understanding of the EU's regulatory design choices in choosing to legislate for AI regulatory sandboxes. The rationale for incorporating AI regulatory sandboxes within the AI Act's ecosystem and the twin goals of fostering innovation and boosting regulatory understanding and compliance are then elucidated. Section 4 considers the intricacies surrounding AI regulatory sandboxes meeting the goal of assisting regulatory compliance while Section 5 critically examines how AI regulatory sandboxes will be able to meet the goal of fostering innovation. This provides the backdrop for an examination in Section 6 of the challenges at play for Member States in operationalising AI regulatory sandboxes. Section 7 moves to identify and probe factors which may have a bearing on whether providers will find AI regulatory sandboxes attractive in practice, including the inability provide temporary regulatory relaxation or to deliver a presumption of conformity to sandbox participants upon exit. Section 8 then moves to provide an original early stage analysis of Member States' preparedness to meet the obligation to establish AI regulatory sandboxes by 2 August 2026 (Art. 57(1), AI Act). Section 9 offers some reflections on what is needed now from Member States and the EU to benefit the roll-out of AI regulatory sandboxes and their situation within other ecosystem supports. The article's conclusion emphasises that the EU must take active leadership steps to ensure cohesion and to guard against the risks of national fragmentation and arbitrage in the operationalisation of AI regulatory sandboxes.

[2] This draws on information in the public domain from Member State authorities and commentary from legal practitioners.

## 2. The Compliance Context

### *2.1 The AI Act's Regulatory Approach*

Given that the AI Act is pan-sectoral in scope and principles-based, compliance is difficult and multi-faceted to get grips with. The AI Act provides an answer to the Collingridge Dilemma whereby waiting too long before regulating technology permits it to become socially and economically embedded, rendering it difficult to regulate later (Collingridge, 1980). However, as a form of early horizontal risk-based regulation that is not rule based, the AI Act suffers from uncertainty of application when combined with technological complexity and innovative planned use cases. The challenge of applying it is greater because the AI Act is the result of a marriage between a well-worn product safety approach now applied to AI, and the new era of what has been branded "digital constitutionalism" (Celeste, 2019) that calls ethical principles and the Charter of Fundamental Rights of the European Union[3] into the regulatory mix. Under the decentralised system which operates for high-risk AI systems, Member States are expected to multi-task, playing the role of innovation facilitator, guide to the AI Act, as well as supervisor and enforcer.

The conformity assessment mechanism for high-risk systems falling under Article 6(2) and Article 6(3), as referred to in Annex III, is a central plank of the AI Act.[4] Notably, from a regulatory design perspective, the AI Act does not follow a public authority evaluation and authorisation approach to high-risk AI systems. Rather it follows the New Legislative Framework approach of pre-existing product safety legislation which revolves around compliance with voluntary harmonised standards and conformity assessment procedures under Article 43 that lead to an EU declaration of conformity under Article 47 and the affixing of a CE marking under Article 48. This approach builds on the New Legislative Framework legislative approach in pre-existing product safety legislation but crucially is designed to consider significant risk of harm to health and safety, and fundamental rights.

[3] Charter of Fundamental Rights of the European Union OJ C 326, 12.10.2012, p.391-407.

[4] These include, for example, AI systems used in HR systems to filter job applicants and AI systems used in university admission decisions. For a detailed unpacking of what qualifies as a high-risk system within the regulatory context of the AI Act see Ruschemeier & Bareis, 2025.

### *2.2 Compliance Obligations on Providers of High-Risk Systems*

Intending providers of high-risk systems constitute the obvious target audience of AI regulatory sandboxes since Article 16(f) of the AI Act requires qualifying systems to undergo an Article 43 conformity assessment procedure before they are placed on the market or deployed. A conformity assessment involves the provider of a high-risk system demonstrating compliance with the mandatory requirements in Section 2 of Chapter III of the AI Act including those associated with trustworthy AI. These span the gamut from accuracy, explainability, transparency to human oversight, cybersecurity and robustness. These requirements provide the baseline within which validation, testing and compliance of high-risk AI system occur, including within AI regulatory sandboxes, the aim being to ensure that there are appropriate data governance practices to guard against biases that could affect health and safety, fundamental rights or lead to illegal forms of discrimination.

Much of the pre-market compliance work involves testing and developing appropriate systems. As prospective providers prepare for market in the EU, they need to develop an Article 9 risk management system for risks to health, safety or fundamental rights. This necessitates the carrying out of appropriate pre-market testing of high-risk AI systems to identify appropriate risk management measures. Giving an equal billing for fundamental rights with product safety considerably complicates the development of such risk management systems (Almada & Petit, 2025; Ruschemeier & Bareis, 2025). Providers also need to put an Article 17-compliant quality management system in place which includes a regulatory compliance strategy and to prepare technical documentation in line with Annex IV that will enable competent authorities and notified bodies to assess compliance. This type of preparing and road-testing is likely to be the bread and butter work conducted in AI regulatory sandboxes.

## 3. Conceptualising AI Regulatory Sandboxes

To understand the appearance of AI regulatory sandboxes in the AI Act necessitates a brief traverse through the meteoric, organic rise of regulatory sandboxes in the fintech space in the last decade as providing the inspiration for the inclusion of an AI regulatory sandbox as a central feature of the decentralised regulatory ecosystem in the AI Act.

### *3.1. The Origin and Spread of Fintech Regulatory Sandboxes*

Regulatory sandboxes form part of the toolkit of anticipatory governance and the mindset that seeks to actively foster innovation and the development of new markets (Ahern, 2025). The OECD defines "regulatory sandboxes" as "controlled experiments with selected participants over defined periods of time" (OECD, 2025, p.8). The brainchild of Sir Mark Walport, the United Kingdom's Chief Scientific Advisor, based on the clinical trials process for pharmaceuticals (UK Government Chief Scientific Adviser, 2015), regulatory sandboxes for fintech first emerged in 2016 for use by the United Kingdom's Financial Conduct Authority to promote competition by allowing selected fintech innovators to engage in controlled testing in a supervised environment guided by the regulator on how to navigate a financial services regulatory landscape not designed with their digital innovation in mind (Ahern, 2019; Miglionico, 2022). A dual role was notable in the FCA regulatory sandbox which was replicated elsewhere. First, "[t]he genius of the regulatory sandbox [lay] in how it provides a sheltered environment to assist FinTech innovators to negotiate the impasse of an unclear regulatory environment while testing the viability of their imaginative products on a scaled-down basis" (Ahern, 2019, p.347). Secondly, the regulatory sandbox's object was to support innovation through promoting competition and market entry (Financial Conduct Authority, 2015, p.5) thus positioning financial regulators in a role that lay far beyond the traditional market gatekeeper.

Regulatory sandboxes have been rapidly diffused by fintech regulators globally and across the EU.[5] In 2024, the author's re aware of regulatory sandboxes for fintech being operational in Austria, Czechia, Cyprus, Denmark, Estonia, Greece, Hungary, Ireland, Italy, Latvia, Lithuania, Luxembourg, Malta, Poland, Portugal, the Netherlands, Slovakia,

[5] Regulatory sandboxes generally operate within an ecosystem of spaces that support innovation such as incubators, accelerator hubs and innovation hubs.

Spain and Sweden, representing some 70 percent of EU Member States. For those Member States who have not joined the trend, factors such as technological expertise, limited human and financial resources may play a part (Raudla et al., 2025). In many cases fintech regulatory sandboxes operating in Member States do not have a statutory footing, or in some cases, transparent set of operating rules, leaving a large amount of discretion as to their operation to relevant regulators.

The fintech sandbox setting typically allows a time-limited, simulated or scaled-down testing to enable the regulator and the innovator “to observe the implications of the impact of the tested innovation [to] draw the appropriate conclusions (OECD, 2025, p.13). Resulting learning around regulatory compliance reduces uncertainty. Sandboxes yield valuable information for both regulators and innovators around whether and how the proposition tested can be put on the market in compliance with existing regulations. For regulators, regulatory sandboxes can deliver invaluable evidence-based regulatory learning (Fenwick et al., 2017) and regulatory learning insights may also arise at a meta level concerning whether regulatory perimeter updating or new bespoke frameworks are needed to meet market needs posed by technological change and new business models (Ahern, 2019, 2021). Regulatory sandboxes may improve the supervisory and risk-assessment capabilities of regulators around new technologies and business models (World Bank, 2025, p.26-28). On the other hand, there is an undeniable tension between the perceived role of regulatory sandboxes as facilitating competition and innovation versus financial services regulators’ expected role in risk mitigation (Ahern, 2019). A particular controversy around fintech regulatory sandboxes which divides commentators is whether and to what extent regulatory rule relaxation should be permitted within a regulatory sandbox with more traditional regulators not open to rule relaxation and others being open to tailor-made sandboxes or being opaque on the issue (Ahern, 2019, 2021; Ringe and Rouf, 2020).

There is a dearth of empirical research on the impact or effectiveness of regulatory sandboxes (Allen, 2025). Scaled-down testing reduces risk to consumers and enables modifications to be made as necessary (Ahern, 2019). There is evidence to support a positive correlation between the introduction of a regulatory sandbox and increased fintech venture investment inflows (Cornelli et al., 2024; Goo & Heo, 2020,). This suggests that a potential reduction in regulatory risk is well received by the venture capital market. In terms of success, most obviously they have helped a selective cohort of fintech innovators to expedite their route to market by navigating highly complex financial services regulatory frameworks not conceived with disruptive innovation in mind.

However, globally, policymakers have reported mixed findings on the question of whether regulatory sandboxes have increased competition in relevant markets (World Bank, 2020, p.35) . Indeed, it is well understood that regulatory sandboxes only directly aid actual participants in the sandbox by addressing regulatory opacity as an obstacle to market entry by assisting with delivering a free regulatory steer from the mouth of the regulator. On the other hand, regulatory sandboxes are highly resource-intensive, helping the chosen few, not the many while innovation hubs and other supports may assist vastly greater numbers of small firms around regulatory barriers in a more timely fashion (World Bank, 2020, p.31).

Despite their proliferation, the European Union did not move to provide a harmonised framework for fintech regulatory sandboxes although there was speculation from scholars as to whether and how this could occur (Ringe & Ruof, 2020; Ahern, 2021). By contrast with fintech sandboxes, the AI Act's incarnation of the regulatory sandbox accompanies a bespoke, risk-based legal framework for AI albeit one whose application in practice raises many complexities. Thus in contrast with fintech sandboxes which function fill a regulatory gap, the European Union in the EU Act was determined to make regulatory sandboxes a lasting feature of agile governance to assist developers to get to grips with compliance and thus aid EU market entry.

Regulatory coherence will assume a broader significance for the EU around sandboxes as it is increasingly making use of regulatory sandboxes with the most obvious comparator with the AI Act model being the Cyber-Resilience Act[6] which contemplates cyber resilience regulatory sandboxes being voluntarily established by Member States for controlled testing of products with digital elements "to facilitate their development, design, validation and testing for the purposes of complying" with the Regulation (Cyber-Resilience Act, Art.33(2)).

### *3.2 Unpacking the EU Rationale for AI Regulatory Sandboxes*

Regulatory sandboxes for AI are thinly conceptualised in the AI Act but the wording employed exhibits similar ideals to their fintech forebears. They are designed to provide "a controlled environment that foster innovation and facilitates the development, training, testing and validation of innovative AI systems for a limited time before their being placed on the market or put into service pursuant to a specific sandbox plan agree between the providers or prospective providers and the competent authority" (Art. 57(5)).

[6] Cyber-Resilience Act Regulation (EU) 2024/2847 of the European Parliament and of the Council of 23 October 2024) [2024] OJ L. 20.11/2024.

This is echoed in the definition of an AI regulatory sandbox in Article 3(55) which alludes to this occurring "under regulatory supervision". The AI Act indicates that competent authorities must "allow broad and equal access and keep up with the demand for participation" (Art. 58(2)(b)) while at the same time giving priority access to EU SMEs and start-ups.

A dual mandate for AI regulatory sandboxes of facilitating innovation and regulatory compliance was present from their initial conceptualisation. Regulatory sandboxes were seen by the EU High-Level Expert Group on AI ("AI HLEG") as an "agile policy-making solution" to promote innovation "without creating unacceptable risks" (EU High-Level Expert Group on AI, 2019, para.29.2). AI HLEG envisaged that regulatory sandboxes for AI would be part of the provision of expert testing facilities for AI. However, it also envisaged that AI regulatory sandboxes could build capacity around fundamental rights impact assessments for experimental AI application (EU High-Level Expert Group on AI, 2019, para.29.2). This second aspect matters, not just for public actors, but for innovators who are private actors since the provision of essential services such as financial services and activities such as credit scoring or insurance scoring are captured within the AI Act's frame on fundamental rights. Here, within the work of the AI HLEG, it is possible to discern the origin of the AI Act's dichotomous conception of regulatory sandboxes for AI as having two quite distinct functions: (i) promoting the development and testing of innovation to expedite market readiness; and as (ii) a supervised means of reducing regulatory risk. This dichotomy of purpose was taken up in the Explanatory Memorandum to the European Commission's Proposal for the AI Act which envisaged regulatory sandboxes as playing a role in supporting innovation and reducing "regulatory burden" (European Commission, 2021a, 1.1). Ultimately, this approach continued to hold sway and in the adopted AI Act, the substantive provisions dealing with AI regulatory sandboxes are housed in Chapter VI, “Measures in Support of Innovation”.

The AI Act’s inclusion of AI regulatory sandboxes within its architecture speaks to an ever-present push-pull tension that exists between regulating a new industry and supporting it (Ahern, 2023). In the context of fintech regulatory sandboxes aims, these tensions are well-rehearsed along with the challenge for sandbox authorities of balancing regulation and innovation objectives (Bromberg et al., 2017; Ringe & Ruof, 2020). A choice between digital regulation and innovation is perhaps a false dichotomy (Bradford, 2024. Nonetheless within the context of rolling out a single market for high-risk AI systems, the success and compatibility of these two aims – compliance assistance and fostering innovation - depends on a coherent conceptualisation of how they work together and play out in the AI Act itself and in the design and operationalisation of the network of AI regulatory sandboxes in the Member States.

## 4. The Goal of Assisting with Legal Regulatory Compliance for High-Risk AI Systems

### *4.1 Regulatory Compliance in the Context of the AI Act*

The role of AI regulatory sandboxes in facilitating regulatory compliance meets the requirement for providers of high-risk systems to put in place a quality management system that details the firm’s strategy for regulatory compliance (Article 17(1)(a) AI Act). AI regulatory sandboxes will be able to assist participants with producing the technical guides required by the AI Act. This is specifically envisaged for the Spanish AI regulatory sandbox. Furthermore, the lessons learned will be more widely diffused. There is an intention is to publicly publish the technical guides prepared within the sandbox to benefit the wider ecosystem (La Moncloa, 2025). The intention is also that the results of the first cohort's participation in the sandbox will be used to inform a report on best practice (La Moncloa, 2025). That indicates smart regulatory thinking around ensuring diffusion of regulatory learning spillover outcomes for this sandbox.

Assessing the robustness of the quality management system will mean engaging with the technical documentation for the product in relation to its design and operation as well as the methodology used on the training datasets and the nature and quality of these datasets. Testing and validation could make systems both more technically robust and more legally robust. One possible beneficial outcome from testing is that it may highlight bias in datasets and models that could lead to discriminatory outcomes; where detected this can then be mitigated. This could, for example, have relevance in relation mitigating feedback loops being skewed by biased outputs affecting future operations. Greater clarity around how the models operate may lead to greater accuracy, explainability, accountability and transparency. In short, adaptive refinement of AI systems is possible.

Compliance here could run the gamut of ethical norms, legal norms and risk-management expectations contemplated by the AI Act. Regulatory sandboxes for AI could also potentially help to deal with the mammoth transverse effects of AI and enable tools to be developed for testing compliance with the gamut of legal risks. Contributing to understanding these interlinkages could increase trust in AI (OECD, 2023, p.12). Competent authorities setting up regulatory sandboxes will provide guidance to sandbox participants on identifying potential risks to health and safety but also to fundamental rights and how to mitigate them and ensure compliance. That will be particularly demanding for Member States as fundamental rights are wide-ranging and values-based and the application of them in practice is not always easy to envisage in advance. It is justifiable that sandbox participation should be halted where significant risk to health and safety or fundamental rights has been identified for which no adequate mitigation has been found (Art.57(11) as in these circumstances firms will understandably have to go back to the drawing board.

### ***4.2 Rule Relaxation and AI Regulatory Sandboxes***

Regulatory derogations have proved a controversial subject for fintech sandboxes with a sharp division between sandboxes that signal a willingness to relax rules on a case-by-case basis and those that are silent on this (Ahern, 2019). Clear regulatory design choices matter particularly here.

The AI Act is not as clear on the question of the application of the AI Act during the sandbox as it could be. Sandbox participants are intending providers on the EU market of AI systems that are entering the controlled testing environment under regulatory supervision “for a limited time before their being placed on the market or put into service” (Art.57(5)). It would therefore seem plausible that provisions of the AI Act that concern AI systems being put on the market or into service are not applicable during sandbox participation although this is not expressly spelt out. However, the better view seems to be that relevant provisions of the AI Act still apply, a view that is confirmed by Article 57(12)'s insulation from the imposition of administrative fines under the AI Act for good faith participants in the sandbox  (but not from breach of the law itself). This contrasts with the position where a national AI regulatory sandbox is set up to include real-world testing (which is optional under the AI Act) where there is an express confirmation that there is no risk of breaches of the AI Act being deemed to occur during real-world testing as it is not treated as placing the AI system "on the market or into service" (Art.3(56)).

As regards other laws, the AI Act is clear that other generally applicable legal frameworks continue to apply within AI regulatory sandboxes, something which will strike some as not sufficiently accommodating to innovators. The AI Act expressly operates without prejudice to other EU law on consumer protection and product safety (Art. 2(9)). Article 57(11) clarifies that the existence of AI regulatory sandboxes "shall not affect the supervisory and corrective powers of the competent authorities supervising the sandboxes, including at regional or local level." In addition, sandbox participation is indicated in Article 57(1) to be without prejudice to the powers of competent authorities associated with the operation of the sandbox such as those operating in the sphere of data protection, equality, consumer protection and competition law (see European Commission, 2021, para.5.4.5). This on its face raises concerns regarding the potential for infringement and enforcement action during sandboxing. In practice it is to be hoped that supervisory authorities act sensibly. The AI Act does give a cue that where discretion is built into the relevant law, national competent authorities should use their discretionary power in relation to AI regulatory sandbox projects in such a way that supports innovation in the EU (Article 57(11) AI Act). The articulation could be clearer, but it would seem to mean, for example, that where legislation permits proportionate application to SME actors, for example, that discretion could be invoked. However, there are no guarantees on that front and national and EU authorities have full remit to supervise and to enforce the law against AI sandbox participants.

For many, the biggest worry arises in relation to how innovators must tackle the hurdle of data use for AI while navigating laws that were not designed with this in mind. There is a very limited exception provided in the AI Act for the processing of personal data within an AI regulatory sandbox for certain public interest AI systems provided that conditions are met (Art.59 AI Act). It is believed that availing of this exemption could be quite burdensome and therefore may not be calibrated to attract relevant parties to apply to an AI sandbox.

This question of regulatory relaxation has been a vexed issue globally for fintech regulatory sandbox operators and one which goes to the heart of the competitiveness of their regulatory offering. Interestingly, the OECD's understanding has been that waiver of relevant legal provisions may occur within a regulatory sandbox for AI (OECD, 2023, pp.13-14.) And in other contexts, the EU has been willing to consider a more flexible stance (European Commission, 2023a). For instance, in the EU's proposal for regulatory sandboxes for medicinal products, a temporary, adapted regulatory framework for developing medicinal products may be provided (European Commission, 2021b). However, the context and rationale here differ from AI regulatory sandboxes, in that where permitted, regulatory sandboxes for pharmaceuticals would most likely be framed to (i) assist in the early stages of novel medicinal product development, and (ii) contribute to the development of new regulatory frameworks. Nonetheless the policy issue does not go away. The influential Draghi Report argued for regulatory sandboxes for clean technologies that would allow temporary rule relaxation while testing (Draghi, 2024). This is complex but arguably it is appropriate to distinguish between regulatory sandboxes that allow for rule relaxation and experimentation while attempting to devise a new regulatory regime, as seen in the EU in the energy sector (Gangale et al., 2023) versus AI regulatory sandboxes under the AI Act where a bespoke regulatory regime is already in place and the focus is on navigating the complexity of its application in practice.

What we can take from the above discussion is that diverging EU policy and regulatory design choices around rule relaxation, exemptions and supervisory powers within EU-devised policymaking for regulatory sandboxes in different settings need a transparent and convincing justification. From a regulatory design perspective, inconsistencies give an impression of normative incoherence arising from an á la carte or siloed approach to conceptualising regulatory sandboxes at EU level.

### *4.3 Potential Compliance Benefits of Sandbox Participation*

AI regulatory sandbox participation may assist with evaluating the efficacy of regulatory compliance efforts. However, competent authorities operating AI regulatory sandboxes are not empowered to deliver certification providing a presumption of conformity under the AI Act. Strikingly, the most an AI sandbox participant can tangibly leave with is a sandbox exit report from the sandbox authority setting out the sandbox activities conducted and related results and lessons learned (Art.57(7) AI Act).

Any resulting acceleration of conformity assessment could be beneficial to achieving compliance before market entry. Sandbox exit reports “shall be taken positively into account by market surveillance authorities and notified bodies, with a view to accelerating conformity assessment procedures to a reasonable extent” (Article 57(7) AI Act). This still falls short of granting a presumption of conformity, particularly as the independence of notified bodies in carrying out conformity assessments is legislatively enshrined (Art.31). On one view, extending a presumption of conformity on sandbox exit could generate safety concerns. This has been a wider concern for sandboxes in general with concern being expressed that fintech regulatory sandbox participation could involve risk-washing and create “a false perception of safety and compliance in the market” (Lezgioğlu Özer, 2024. See also Brown & Piroska, 2022).

However, an alternative perspective is that if the provisions were strengthened to offer the ability to obtain a presumption of conformity on sandbox exit would prevent duplication of effort and delays en route to market. In line with this, it has been suggested that a presumption of conformity would be helpful to operationalise “genuinely simplified and accelerated conformity assessments explicitly linked to sandbox exit documentation” (Novelli et al., 2025, p.30). As matters stand, all need not be lost in practice if intelligent workarounds are employed by AI regulatory sandboxes. For instance, a pragmatic approach is being adopted for Luxembourg’s AI regulatory sandbox. It is said that Luxembourg’s AI sandbox will “not provide certification [but] will provide an assurance of compliance .... Notified bodies [who conduct conformity assessments] will provide the stamp” (Labro, 2025).

Although participation in an AI regulatory sandbox will not deliver a regulatory approval or presumption of conformity, indirect regulatory compliance benefits are likely to accrue from sandbox participation. As regulatory requirements constitute a major barrier to entry to the single market, the opportunity for regulatory dialogue and mutual learning for regulators and innovators mark out regulatory sandboxes as highly valuable for both innovators and regulators (Ringe & Rouf, 2020).

Indeed, the chance for firms to break down barriers with regulators may outweigh the disadvantages of any perceived need for firms to 'educate' regulators on how their AI systems function. Early regulatory engagement within an AI regulatory sandbox could be beneficial for AI firms seeking to strategically capture market share where regulatory responses around AI systems and risk management have not yet fully coalesced (Frederiks et al., 2024). Furthermore, providers may be attracted to sandboxes precisely because in seeking to be AI Act-compliant they worry that they may miss a risk that would retrospectively be deemed to be foreseeable (Schuett, 2024).

## 5. The Goal of Fostering Innovation and Competitiveness

### *5.1. The Nature of this Goal*

One of the puzzling aspects of the refashioning of regulatory sandboxes for use in the AI Act has been deciphering the practical import of quixotic references during the Regulation's legislative journey to their role in fostering innovation. It is noteworthy that the definition of an AI regulatory sandbox (Art. 57(5) AI Act) expressly focuses on facilitating innovation to get to market by providing "a controlled environment that fosters innovation and facilitates the development, training, testing and validation of innovative AI systems ... before their being placed on the market or put into service....”[7] This is joined by the AI Act’s broader expressed meta objective that AI regulatory sandboxes will contribute to “competitiveness and facilitating the development of an AI ecosystem” (Art. 57(9)(b), (e) AI Act). These, on their face, are laudable objectives but the failure to adequately separate out innovation-related goals from regulatory compliance goals is unsatisfactory.

The first aspect in which impact on innovation arises is at the point of assessment of selection for admission. ‘Innovativeness’ frequently operates as a central express or de facto filtering criterion for selection for entry to fintech regulatory sandboxes (Ahern, 2019). This puts the regulator in the challenging position of adjudicating on innovativeness in determining which firms should be accepted to a limited entry sandbox, a task which manifestly lies outside regulators’ traditional comfort zone and skillset. Ranking applicants on innovativeness thus typically leads to the charge that regulators should not be ‘picking winners’ (Knight & Mitchell, 2019, p.14). Interestingly, in the first call for the Spanish AI Regulatory Sandbox pilot specified that the degree of innovation or technological complexity of the product or service was weighted at 20 percent in the selection evaluation criteria.

The extent to which real experimentation and innovation will occur within, and be enabled by, AI regulatory sandboxes remains to be seen: once their design and operation is realised along with a throughput of sandbox admissions and exits, evidence will become available for future study. The facilitation of innovation in the truest sense may be rather limited because of the degree of technological readiness that innovation will need to be at to benefit from a sandbox. AI regulatory sandboxes should not support AI systems at low technology readiness level even where they appear to hold promise as they are more suited to a lab environment or testbed. Indeed, the OECD suggests that, in general, regulatory sandboxes are most useful for Technology Readiness Levels 7-9 rather than early-stage innovations (OECD, 2025, p.21). If national AI regulatory sandboxes are operated to prefer admission of systems that are close to ready to enter the market, the focus of the sandbox is likely to be on pre-marketing compliance and risk mitigation over early stage experimentation that would perhaps be more appropriately trialled outside AI regulatory sandboxes, such as in real-world testing and experimentation facilities which permit sector-specific testing and refinement.

[7] See also Art. 3(55) AI Act.

As such, the likely contribution to innovation of an AI regulatory sandbox is often likely to be indirect in the form of assistance with navigating regulatory interfaces en route to market. On this view, testing and validation within a regulatory sandbox should occur in service of regulatory and ethical compliance. To elevate the contribution to innovation may risk mission confusion for regulators administering AI regulatory sandboxes, something which is best avoided (Genicot, 2024, p.3).

That said, AI regulatory sandboxes could assist start-ups by providing access to testbeds and other valuable resources such as compute power or synthetic datasets (particularly useful to avoid the obligations associated with using personal data). National sandboxes that choose to do so may potentially meaningfully contribute to innovation facilitation and increase their sandbox's relative competitiveness measured in terms of their attractiveness to applications from start-ups and SMEs.

### 5.2 The Contribution of Testing and Validation

The controlled environment within the sandbox itself is where "development, training, testing and validation" will be facilitated while following an agreed sandbox plan (Art. 57(5) AI Act).

#### *5.2.1 Testing*

As regards testing methods, scant detail is available in the AI Act itself beyond the specification that testing may take the form of a supervised simulation or alternatively, limited scaled-down real-world testing may be carried out by providers with the participation of a user sample (Art. 57(5) AI Act), an option added by the Council during negotiations on the AI Act (Council of the European Union, 2022). This would usefully allow a provider or prospective provider to test in cooperation with deployers or prospective deployers so as to provide a real-world testing under supervision. Thus user insights from supervised real-world testing may help to inform practical refinements by developers to their AI systems including around ethical implications and responsible innovation insights.

#### *5.2.2 Validation*

Art.57(5)'s allusion to validation indicates that AI regulatory sandboxes could be configured to offer secure pre-marketing and pre-deployment validation of AI systems allowing testing and evaluation of the functionality of AI systems for technical issues that are separate from regulatory concerns. This is not assisting innovation in terms of early stage ideation but rather innovation in terms of close to market ready technical validation. A sandbox environment could enable the robustness of an AI system to be stress-tested outside of live testing within a controlled, monitored environment that mimics the real world. This could be considered a contribution to fostering innovation given that such testing could be helpful for ensuring stability, security and scalability of AI systems.

As a controlled and secure environment with an ethical and legal container, an AI regulatory sandbox will also enable an AI system to be tested. From the definition of "testing data" in Article 3(32) testing here refers to data used in the context of evaluating an AI system's performance against its expected performance before it is put on the market or into service. It may also relate to testing its conformity with the requirements of the AI Act. Evaluation of a trained AI system while monitoring it in an isolated environment that will not impact on live operations in the real world minimises risk of harm including data breaches.

Within the sandbox environment, developers could run simulations and, where necessary, iterate and fine-tune algorithms. Bugs in the system may also be identified, making the AI system more reliable. Controlled failure in this safe environment could prompt creative thinking around responses and improvements, with the opportunity to mitigate and manage risks before an AI system is put on the market or into service.

## 6. Challenges for Member States in Designing and Operationalising AI Regulatory Sandboxes

Difficulties stem from the AI Act's very broadly expressed objectives around AI regulatory sandboxes (Article 57 and 58) and the lack of instructional precision for Member States concerning actualising the set up and operation of these sandboxes (Buocz et al., 2023)

The logistical and design challenge presented by Article 57(1)'s ambitious deadline for establishment of AI regulatory sandboxes of 2 August 2026 will undoubtedly be exacerbated by the level of discretion given to Member States in operationalising their sandboxes when combined with a knowledge deficit around AI regulatory sandboxes as a new type of regulatory beast while coordinated EU level guidance is awaited. Therefore, as elaborated upon below, there is real potential for quite divergent approaches and priorities being taken by Member States as they make regulatory design choices around the establishment and operation of their AI regulatory sandboxes.

### *6.1 Resourcing and Capacity as Design Choices for Member States*

National competent authorities establishing sandboxes and other relevant regulators undoubtedly will be on a very steep learning curve. The capacity-building challenges are formidable as competent authorities are expected to bolt together and apply technical understanding, regulatory understanding and fundamental rights expertise to complex use cases. It is well understood that the allocation of regulatory resources is a regulatory design choice (Aviram, 2012) and one that will play heavily on governmental minds as they work out the institutional design for AI regulatory sandboxes as well as national market surveillance authorities. To function optimally, regulatory sandboxes will need, not only the right infrastructure but also a diverse range of appropriately skilled personnel across the gamut of potential opportunities and risks that may arise. The extent to which AI sandboxes can assist prospective AI providers to navigate regulatory uncertainty within the testbed and deliver trusted, useful insights that can be actioned in line with the obligations on providers of high-risk systems will be correlated to the level of cross-cutting expertise available. Fintech regulatory sandboxes have proven to be extremely resource intensive given the hands on supervision needed and AI regulatory sandboxes will be no different. AI regulatory sandboxes also call for technical expertise across health and safety as well as broad-spanning ethical and legal expertise including on fundamental rights. There will be an inescapable need for sandboxes to develop what Almada and Petit term “technological literacy” (Almada & Petit, 2025, p.112). Without personnel who can appropriately grasp the technical aspects of the AI systems before them, a sandbox may lose trust (Alaassar, et. al, 2020). Market perception of a particular AI regulatory sandbox’s resourcing and expertise, as relevant aspects of its design choices, is likely to be linked with market reception and relative perceived attractiveness in terms of responsiveness to market demand for participation.

To further complicate matters, AI regulatory sandboxes ideally need to have appropriate expertise to call upon to be able to coordinate effectively with regulators who enforce law that high risk AI systems may touch upon. Regulatory cooperation and coordination among multiple regulators in fintech sandboxes are known to be challenging (Plato-Shinar & Godwin, 2025). With AI regulatory sandboxes under the AI Act, there needs to be an understanding of how other regulators’ terrain may be triggered as sandbox participation is without prejudice to the powers of other supervisory authorities including in the sphere of data protection, equality, consumer protection and competition law (Art.57(11) AI Act). AI sandbox operators will therefore need to cooperate with and involve relevant national agencies including those overseeing fundamental rights protection.

### *6.2. Implementation and Co-ordination Challenges for Member States*

With the detailed arrangements for operationalising regulatory sandboxes not being fleshed out in the AI Act, in accordance with Article 58, the Commission is expected to produce detailed arrangements in an implementing act for the operationalisation of AI regulatory sandboxes. The approach of leaving detailed rules to be determined in implementing acts or guidance is a well-worn feature of contemporary EU legislation. But where the timescale for issuing such guidance proves badly matched to implementation timelines, problems around consistency and coherence are likely to prove inevitable as Member States are forced to act in a vacuum.

Thus while the AI Act is billed as a maximum harmonisation measure, as regards high-risk applications, it could see both gold-plating and under-plating in its operationalisation on the ground by Member States bent on moving forward with AI regulatory sandboxes and other supports to secure competitive advantage while lacking concrete guidance around interpreting the strictures of the AI Act in practice. This is a significant risk arising from regulatory design choices coupled with overly ambitious implementation deadlines for both Member States and the roll-out and work of the AI Office and AI Board as well as that of the standardisation bodies.

### *6.3 Risks of Fragmentation and Arbitrage within AI Regulatory Sandboxes*

While a centralised sandbox might have been operationally difficult to achieve, there are some would have preferred a centralised enforcement model to navigate the complexities presented by the AI Act (Stahl et al., 2022). The EU could alternatively have opted solely for pan-EU sectorial sandboxes which would have aided uniformity.

With the decentralised approach taken, regulatory sandboxes will only work effectively if concerted efforts are made to ensure pan-EU coordination around their operation so that there is a consistency of approach and shared mutual understanding around relevant learnings by the sandbox authorities. Otherwise providers will exploit this to play sandbox arbitrage in choosing which sandbox to target. As is evident with the GDPR (Ebers et al., 2021; European Parliament, 2021, p.3;), differences of approach at national level to the AI Act could jeopardise the rule of law (Söderlund and Larsson, 2024) Uneven approaches within national sandboxes could contribute to this. Thus, there is a need to "reduce cross-border dissonance" to avoid fragmentation and arbitrage (Ahern, 2021, p.428).

The potential for divergent national approaches to AI regulatory sandboxes giving rise to fragmentation in the interpretation and operationalisation of the AI Act has not gone unnoticed (Lezgioğlu Özer, 2024). The detrimental effects for legal certainty for AI innovators of differing approaches among sandboxes to core compliance issues have also been highlighted (Tech Tonic, 2025). The Commission for its part has acknowledged the need for both Member States and the AI Office to "step up their efforts to facilitate a smooth and predictable application of the AI Act" (European Commission, 2025b, p.21).

Not only could national AI regulatory sandboxes differ in their approaches, a parallel challenge concerns the possibility that interpretation by authorities within AI regulatory sandboxes of the AI Act in relation to their application in practice could potentially be at odds with the approach taken in the standards which will assume quasi-legal standing. The potential for friction may diminish where national standards bodies are involved in both conformity assessments and establishing AI regulatory sandboxes. The Czech

Standards Agency has been tasked with responsibility for creating the Czech AI regulatory sandbox (Únmz, 2025).

Matters are complicated by the delay in the draft standards being ready, with the work now expected to continue into 2026, likely giving very little, if any, time for the Commission to review the draft standards, and for sandbox operators to assimilate their impact if approved, before AI sandboxes come into effect. The Commission could adopt common specifications under Article 41 to fill the gap but time is also not on the Commission's side in that regard.

The ball now rests in the Commission's court. The Commission is empowered to set guidelines to avoid fragmentation and to ensure that AI sandbox participation "is mutually and uniformly recognised and carries the same legal effects across the Union" (Art. 58(2)(g) AI Act). Such guidelines will be important both to clarity and coherence of approach and to preventing national fragmentation across the Member States.

Finally, it bears saying that some AI developers may be drawn to sandbox regimes that are perceived as less stringent (European Parliamentary Research Service, 2022, p.5). Arbitrage around differences in regulatory practice is a common game (Pollman, 2019; Pošćić & Martinović, 2022). As Ringe and Rouf (2020, p.611) note, "high regulatory requirements are a common incentive for arbitrage activities". Providers will exploit perceived differences to play AI sandbox arbitrage with the result that some sandboxes may attract far more interest from innovators than others. An AI regulatory sandbox may attract industry interest where asymmetries of information or a willingness to be 'pro-tech' leads to a reputation for regulatory capture or at least an easier journey through a particular sandbox over others (Baldini & Francis, 2024; Yandle, 2024). This highlights how much rests on design choices and appropriate guidance to Member States that balances their autonomy in operational matters with the need for levels of shared approaches. Coordination by Member States through their participation in the AI Board is also vital to ensure that that the effectiveness of AI regulatory sandboxes is not hobbled by inconsistencies.

## 7. How Attractive Will AI Regulatory Sandboxes Be to Innovators?

Participation in an AI regulatory sandbox is optional. As the World Bank has noted in its review of the global impact of regulatory sandboxes, in the quest to reduce compliance choices there are alternatives to sandboxes which "might not be the best tool for reducing costs" (World Bank, 2020, p.32). For the AI Act, unclear messaging could prove a hindrance. One study demonstrated some market confusion about what the term "regulatory sandbox" connotes under the AI Act and that lack of clarity around the goals of AI regulatory sandboxes may deter participation (Lanamäki et al, 2025). That is hardly surprising given the embryonic detail in the AI Act and the early stage of their operational planning. Nonetheless, the nub of the issue is that "[i]n order for a sandbox to be attractive, it must convey a certain benefit to the admitted companies" (Poncibó & Zoboli, 2022, pp.29-30). Or as the OECD notes, "a sandbox must respond to a real demand, it should not be a solution in search of a problem" (OECD, 2025, p.20).

The global experience of fintech regulatory sandboxes has shown that the design characteristics of a sandbox offering influence its effectiveness and its relative attractiveness or competitiveness (Ahern, 2021, 2023). These include entry criteria, time to selection decision and the conditions offered in the sandbox including regulatory flexibility. These types of factors are equally likely to be of relevance for the competitiveness of AI regulatory sandboxes under the AI Act as they compete with each other for business.

### *7.1 Insufficient Trade-Offs for Sandbox Participation?*

A question is whether appropriate trade-offs in favour of innovation have been made in the AI Act to incentivise effective sandbox participation (Ranchordas, 2021). A very real question is whether demand for AI regulatory sandboxes will be affected by potential downsides highlighted area which include (i) no relaxation of applicable laws for most innovators testing within the controlled environment of an AI sandbox (as discussed earlier), and (ii) no ability to deliver a certificate of conformity. This raises a putative demand-side challenge for AI regulatory sandboxes under the AI Act that may deaden the rate of applications. Other relevant factors that may influence innovators' choices are excavated below.

Creating the right atmosphere in regulatory sandbox interactions to build trust and cooperation is crucial for sandbox reputation (Fahy, 2022). Without trust innovators may be sandbox-shy. Innovators jealously guard their intellectual property and may feel disinclined to apply to AI regulatory sandboxes where they may be required to disclose algorithms, trade secrets, and confidential information to competent authorities who are also potentially dealing with their competitors (even if the authorities are subject to confidentiality obligations under Art.78). Reassurance may thus be needed. To address this, the Spanish AI regulatory sandbox rules specified that information from sandbox participants around "commercial plans, industrial and intellectual property rights or business secrets, as well as the data and information facilitated will be subject to a strict regime of permanent confidentiality."[8]

Another potential fear of prospective providers is that participation in an AI regulatory sandbox may reveal that they are far from having their ducks in a row and that when regulators are shown 'under the hood' they may not like what they see. Innovators will be conscious of the power of sandbox authorities to temporarily suspend the testing process to allow mitigation to be worked on or to permanently suspend testing if they reach the view that no effective mitigation is feasible. In either case, the AI Office must be informed (Art.57(11) AI Act). This visibility of potential shortcomings may deter potential sandbox applicants from going down the sandbox route.

As regards regulatory compliance, other supports may appeal including those arising from the obligations on Member States to provide training and guidance tailored to the needs of SMEs including start-ups and to establish channels of communication with innovators that will support their innovation journey through guidance and provide responses to specific queries about the application of the AI Act (Art. 62(1)(b), (c), Art. 70(8) AI Act). AI innovators may also benefit from private sector AI Pact (European Commission, 2023b) activities which involve peer knowledge sharing around best practices. While these supports may complement AI regulatory sandboxes, they will also compete with them. For queries around regulatory compliance, once up and running, a proportion of AI innovators can be expected to flock to the planned centralised AI Act Service Desk within the AI Office as a first port of call to provide regulatory compliance advice to start-ups and SMEs in all EU languages based on questions being submitted via an online platform.

[8] Resolution of December 20, 2024 of the Secretary of State for Digitalization and Artificial Intelligence calling for access to the controlled testing environment for reliable artificial intelligence, Article II.4.8.

Well-resourced innovators may prefer to opt to carry out their own independent, unsupervised real world testing of AI systems under predefined conditions (Art. 60(4) AI Act) or to pay to access independent testing and auditing including AI sandboxes that are privately provided by Sandbox-as-a-Service providers who provide a similar facility but do not directly expose AI innovators to the scrutiny of authorities and regulators.

For some, the failure to provide greater protection against liability during sandbox testing may deter sandbox participation. This signals an unfortunate failure to appropriately balance the importance of encouraging participation in order to foster innovation (Lezgioğlu Özer, 2024). Sandbox participants could be exposed to declarations that they have breached applicable law even if they will be insulated from administrative fines for breaching the AI Act provided that they have followed their sandbox plan and guidance from the national competent authority in good faith (Article 57(12) AI Act). However, there is still the potential for civil liability actions by third parties (under national or EU law) arising out of participating in an AI sandbox. This encompasses liability under national and EU law for harm to third parties that occurs during controlled testing within the sandbox that lies with the innovator (Article 57(12) AI Act). The approach in Recital 141 of the AI Act of seeking informed consent of natural persons to participate in testing under real world conditions is sensible but does not adequately dispose of liability concerns. Such concerns could conceivably be addressed in the Commission's future detailed rules on the operation of AI regulatory sandboxes in relation to the terms and conditions applicable to participants. Insurance is a common method of diffusing risk in a commercial context. It could be sensible to specify that firms participating in AI sandbox participants must take out insurance cover that would indemnify third parties for loss or harm incurred as a consequence of the sandbox experimentation. Indeed, insurance was the solution to liability issues arrived at for Autonomous Vehicles in the United Kingdom in the Automated and Electric Vehicles Act 2018 (Morgan, 2025). The issue assumes more relevance with the withdrawal by the Commission in 2025 of the proposed AI Liability Directive (European Commission, 2022b).

### *7.2 Direct Engagement with Standards and Conformity Assessments*

Providers of high-risk AI systems need to have their systems undergo a pre-market conformity assessment procedure under Article 43 demonstrating compliance with the mandatory requirements in the AI Act. Providers can go to a notified body to demonstrate compliance or alternatively once harmonised standards exist and are applied by the provider, the provider can use the internal control procedure under Annex VI without the need to involve a notified body (Art.43(a) AI Act). Where high-risk systems conform with harmonised standards, they benefit from a presumption of conformity with relevant obligations under the AI Act (Art. 40(1) AI Act).

Thus, in contrast to participation in an AI regulatory sandbox, direct engagement with and application of standards under the AI Act provides a direct compliance pathway under the AI Act. Standardisation is a bottom-up process led by industry bodies envisaged by the AI Act as providing technical solutions to ensure compliance and to promote "legal certainty" (Art.40(3) AI Act; Golpayegani et al., 2023). These standards will be integral to practical compliance. However, drafting these standards is a complex task involving a collaborative process that remains protracted and ongoing. Indeed, there are other examples of the European Standardisation Organisations struggling to deliver harmonised standards requested by the Commission in a timely fashion (Tartaro & Panai, 2025). When they arrive, harmonised standards from the European Standards Organisations will form an obvious starting point for compliance for AI developers wanting to put AI systems on the market or into use in the European Union. In other EU product safety contexts, it has been noted that the presumption of conformity that attaches to harmonised standards "renders any other means of achieving compliance more costly and time-consuming" (Gornet and Maxwell, 2024, p.10). Standards will therefore likely provide the central conduit within quality management systems for demonstrating that high-risk systems comply with the AI Act. As such, it has been cogently argued that standardisation is "where the real rule-making" for the AI Act will take place (Veale & Zuiderveen Borgesius, 2021, p.105). Importantly, it has already been judicially established that harmonised standards "form part of EU law".[9] However, any potential attractiveness of direct engagement with standards over sandbox participation begs the question of these standards being finalised and put in place.

Some anecdotal evidence from those providing regulatory advice  on the AI Act suggests that outside the category of small, start-ups, larger well-resourced firms who can afford competent advice and who are strong on testing and validation may feel confident to rely on legal advice and conformity assessment without engaging with a sandbox route that would potentially prolong and duplicate the process. On the other hand, for start-ups and SMEs, the alternative of grappling with standards as a route to a presumption of conformity may bring its own challenges in the absence of access to relevant expertise and testing which can prove costly (Tartaro & Panai, 2025, p.126).

---

[9] *James Elliot Construction Ltd v Irish Asphalt*, C-613/14, 2016 https://eur-lex.europa.eu/legal-content/EN/TXT/?uri=CELEX%3A62014CA0613, para.40; Public.*Resource.Org Inc v European Commission* CJEU Case C-58, 8/21, 2024 https://eur-lex.europa.eu/legal-content/EN/TXT/PDF/?uri=CELEX:62021CJ0588, para.61.

A conundrum does arise concerning whether standards will be able to meaningfully engage with the amorphous nature of fundamental rights concerns. This is open to serious question (Castets-Renard & Besse, 2023; Giovannini, 2021; Gornet & Maxwell, 2024). There is consequently a possibility that self-assessment of compliance with technical standards while bypassing regulatory sandboxes could lead to an element of risk-washing or ethics-washing as innovators focus on finding the lowest common denominator that meets the relevant metric in the race to market (Gornet & Maxwell, 2024).

## 8. Member States' Preparedness for AI Regulatory Sandboxes

As the work on standardisation lags behind, the following analysis of the status quo on AI regulatory sandboxes is based on analysis of public domain discussion of Member States' implementation plans. It divides preparedness into four categories: (i) AI regulatory sandbox up and running; (ii) implementation plans actively underway; (iii) tangible intentions declared; and (iv) detailed plans not publicly in evidence.

Stakeholder consultation and engagement at Member State level is smart in order to assist with buy-in for the AI regulatory sandbox model design adopted (OECD, 2025, p.21) but being able to devote sufficient time for this is difficult when preparing to meet an ambitious implementation timetable that will often involve drafting and adopting national legislation as well as operationalising the sandbox. At the time of writing there is just under a year to go to the 2 August 2026 deadline and for Member States the challenge is to not just to make decisive decisions around structure and resourcing but to get primary legislation drafted and adopted in good time to have the AI sandbox up and running.

### *8.1 AI Regulatory Sandbox Up and Running*

Spain has in place national AI regulatory sandbox legislation[10] and sandbox guidelines.[11] Its establishment of a single all powerful supervisory AI agency (AESIA) with a built-in sandbox is aimed at making it easy for businesses to navigate compliance. A dedicated AI Act sandbox pilot opened in late 2024 based on the notion of a pilot AI regulatory sandbox already conceived in conjunction with the European Commission (European Commission, 2022a). The Spanish AI regulatory sandbox pilot for high-risk AI systems includes the national data protection agency and the agency that regulates medicines and medical devices. The first call for high-risk AI systems was open from late December 2024 to 30 January 2025. The Spanish AI regulatory sandbox then became open for testing on 3 April 2025 when the first AI Act testing environment was activated. Of the 44 applications received, 12 projects were selected to participate (La Moncloa, 2025).

---

[10] Royal Decree 817/2023.

[11] Resolution of December 20, 2024 of the Secretary of State for Digitalization and Artificial Intelligence calling for access to the controlled testing environment for reliable artificial intelligence. This resolution specifies and develops the provisions necessary to complete the legal framework for the AI regulatory sandbox.

See, further the website for Sandbox IA, https://avance.digital.gob.es/sandbox-IA/Paginas/sandbox-IA.aspx.

Official sources indicate that the selection process aimed to include a diversity of high-risk AI systems within this sandbox pilot so as to contribute as widely as possible to regulatory learning by assisting a range of sectors in gaining greater understanding (La Moncloa, 2025). In gaining an understanding of the breadth of projects that regulatory sandboxes will potentially deal with it, it is interesting to observe the high-risk purposes of this first cohort of participants. The projects selected were for high-risk AI systems in the area of access to essential public or private services (a credit rating, solvency analysis and credit risk tool, and an AI system for transcribing emergency communications); biometrics (digital age verification, a facial recognition technology AI system, employment: a HR AI system for checking candidates' applications against job requirements, an AI system for matching freelancers with business projects. Two AI systems selected dealt with critical infrastructure: an early warning system for incidents in crowds, and an AI system for cybersecurity in data flows around electricity distribution. Other applications were in relation to machinery and health products including an AI-managed telephone system for monitoring clinical patients (La Moncloa, 2025).

### *8.2 Implementation Plans Actively Underway*

While Spain is out in front in terms of preparations, a number of other Member States have made decisions on how they want to build out the AI regulatory sandbox and are advancing further in laying the groundwork for the journey to implementation. Within this category we find Lithuania, Latvia, Poland, Croatia and Finland. Here we find representation from Baltic states are strong on fintech, AI and digitalisation and have been responsive to crafting the AI regulatory sandbox mandate. Lithuania, a Member State, that has a strong reputation for regulatory agility in the fintech sphere, announced in October 2024 that it planned to move ahead with creating an AI regulatory sandbox with the intention of speeding up the implementation of the AI Act on the ground (Ministry of the Economy and Innovation of the Republic of Lithuania, 2024). The sandbox will be run by Seimas, the Innovation Agency and Communications Regulatory Authority and its rules were amended in early 2025 to accelerate this. Meanwhile Latvia is planning an AI regulatory sandbox that would sit under its National AI Centre and passed an enabling law in 2025.[12] Other Member States are on the journey to getting national law amended to enable an AI regulatory sandbox. Of these, Poland has been solidly working to put the building blocks in place. Following a public consultation in late 2024, in February 2025 Poland's Ministry of Digital Affairs revised its draft Artificial Intelligence Act that when adopted will deal with the establishment of AI regulatory sandboxes (Ministerstwo Cyfryzacji, 2025). Croatia and Finland, are in the process of drafting legislation to regulate AI and create an AI regulatory sandbox. Finland's draft proposal on AI regulatory sandboxes was the subject of a public consultation and the rules are expected to be in force by 1 February 2026, some six months prior to the EU deadline for establishing an AI regulatory sandbox. The sandbox will be run by Traficom, the Finnish Transport and Communications Agency.[13]

### *8.3 Tangible Intentions Declared*

A further cohort of four Member States have declared tangible intentions around the planned structuring of the sandbox. Hungary has said that it will establish a dedicated AI Office to operate as a 'one stop shop' on AI and to provide an AI regulatory sandbox for controlled testing (Petrányi et. al, 2024). Luxembourg has committed to legislating to establish a regulatory sandbox for AI (The Government of the Grand Duchy of Luxembourg, 2025). The Netherlands and Czechia have developed plans for the implementation of an AI regulatory sandbox.

---

[12] https://labsoflatvia.com/en/news/artificial-intelligence-centre-to-oversee-ai-development-in-latvia
[13] https://www.hannessnellman.com/news-and-views/blog/eu-ai-act-latest-regulatory-developments-in-finland/

### *8.4 Detailed Plans Not Publicly in Evidence*

16 Member States or 59 percent of the bloc (Austria, Belgium, Bulgaria, Cyprus, Denmark, Estonia, France, Germany, Ireland, Italy, Malta, Portugal, Romania, Slovakia, Slovenia and Sweden) have not publicly announced detailed plans and may be at varying stages of advancement in scoping how to create regulatory structures under the AI Act while awaiting further guidance on best practice from the AI Board's working group on AI regulatory sandboxes on the roll-out of sandboxes.

### *8.5 Analysis*

Member States differ markedly in terms of levels of preparedness for implementing a national regulatory sandbox. Early movers among the Member States have expressed keenness to leverage the competitive advantage that moving ahead affords companies and countries in gearing up for the AI Act.[14] Of these, Spain has been most proactive and is the only Member States that falls into category (i) AI regulatory sandbox up and running. Five Member States (18.5 percent)_ were found to fall into category (ii) Implementation plans actively underway. A further four Member States (14.8 percent) were characterised as within catgeory (iii) Tangible intentions declared. The majority of Member States, 16 Member States or 59 percent of the bloc, do not currently have detailed plans not publicly in evidence which is a matter of concern given the intricacies involved in regulatory design, recruitment and training.

[14] Some backtracking may be needed once the Commission and the AI Board formally firm up on the parameters for AI regulatory sandboxes, thus leading to early initiatives being simply dubbed as pilot AI regulatory sandboxes.

Notably, of the Member States who have already taken implementation steps or have clear plans for AI regulatory sandboxes (categories (i) to (iii)), seven of the nine Member States have pre-existing experience of running fintech sandboxes (the exception being Croatia and Finland). This indicates that prior national experience running a fintech regulatory sandbox may have proved advantageous, providing some transferable design experience that may have helped to expedite planning. On the other hand, the 16 Member States who do not have publicly announced detailed planning are evenly divided between countries with pre-existing fintech regulatory sandboxes (Austria, Cyprus, Denmark, Ireland, Italy, Malta, Portugal and Slovakia) and those without (Belgium, Bulgaria. Estonia, France, Germany, Romania, Slovenia and Sweden), perhaps reflecting the multi-pronged regulatory and resourcing challenges of responding to what is required of them under the AI Act.

In terms of design choices, existing regulatory sandboxes in Member States could be adapted to serve as AI regulatory sandboxes for the purposes of the AI Act. For instance, Malta's Digital Innovation Authority has a Technology Assurance Sandbox, in place since 2022, would seem well placed to cover the requirements of the AI Act. National data protection authorities in Luxembourg, Denmark and France have valuable experience of running regulatory sandboxes where AI intersects with the GDPR.[15]

[15] General Data Protection Regulation (EU) 2016/679 (as amended) OJ L119, 04.05.2016; cor. OJ L 127, 23.5.2018.

The option also exists for Member States to partner with each other to offer cross-border sandboxes (Art. 57(2) AI Act). This pooling of resources could be particularly helpful to Member States who have less resources available to them or who have little or no experience of running regulatory sandboxes. That said, reflecting the additional complexities at play, there is little in the way of positive existing precedent for multi-jurisdictional fintech sandboxes (Pošćić & Martinović, 2022) even though they have sometimes been considered with a view to enabling scaling up (Ahern 2021; Thomas, 2018; Truby, et. al, 2022). So far there are no indications of this option being pursued.

Inter-jurisdictional regulatory competition between regulators may be considered healthy and potentially ultimately welfare enhancing (Esty, 2000; Hertig, 2000). The same may be said for sandbox environments. The current early gestational stage of AI regulatory sandboxes in the EU does not lend itself to a meaningful comparative analysis of regulatory design choices. However, the prognosis is that decentralised regulatory sandboxes that distinguish themselves as efficient and helpful to navigating the AI Act and adjacent regulatory frameworks may hold value in the eyes of innovators as regulatory costs to market entry present as an obstacle to market entry. Member States will likely be judged on the agility and competitiveness of a given regulatory sandbox as against others. This assessment will depend in part on the wider design choices of Member States in organising their AI regulatory space. For instance, all things being equal, a jurisdiction with a sandbox regulator who will provide a single point of contact (as seen, for example, in Luxembourg's plans) may prove more attractive than a distributed model where innovators must consult with a vast range of sectoral regulators.

## 9. What is Needed Now

### *9.1. Operationalising Member State AI Regulatory Sandboxes*

As this paper has found, some 59 percent of Member States appear to have no detailed publicly announced plans in relation to AI regulatory sandboxes while others are in varying stages of planning. If the EU were to accede to calls to postpone the implementation of some parts of the AI Act, this would give Member States some additional breathing room to get their AI sandboxes operationalised. However, either way, given the complexities of designing, legislating for and operationalising AI regulatory sandboxes, it would be a mistake for Member States to delay actively making implementation plans even if there some adjustments may be needed as centralised EU guidance become available.

### *9.2 EU Implementing Guidelines for AI Regulatory Sandboxes*

It is hard to disagree with the assessment that the final text of the AI Act on regulatory sandboxes did not yield a text "as concise or coherent as many had hoped" (Yordanova, 2024). To avoid the spectre of AI regulatory sandboxes being dismissed as 'hype' (Johnson, 2023; Xiao, 2023) it is imperative that the EU show leadership to guide Member States on their journey towards operationalising EU regulatory sandboxes.

Under Article 66(k) of the AI Act, the AI Board, with representation drawn from all Member States, is tasked with supporting national competent authorities "in the establishment and development of AI regulatory sandboxes". The use of delegated acts to provide guidance allows for regulatory agility. However as the clock is ticking on the AI Act, further delays in agreeing regulatory standards and guidelines for operationalising AI regulatory sandboxes will create uncertainties for both Member States authorities.

In practice, one of the vexed questions that arises is whether or not a given AI system qualifies as high-risk or whether it benefits from the materiality derogation under Article 6(3). Treating this as a threshold issue to address before applying to an AI sandbox ensures that only genuinely high-risk applications participate in an AI regulatory sandbox. Ideally, Member States should be equipped to address this in a cohesive way with the aid of national helpdesks, supports and EU guidance and supports.

This article pointed up the rather torturous treatment around liability issues within sandboxes. As such, the guidelines should be clear and should provide an express confirmation that there is no risk of breaches of the AI Act being deemed to occur during sandbox activities. Member States would also benefit from centralised EU guidance on how to address the intricacies of the interface between AI and the GDPR as highlighted in the Draghi Report (Draghi, 2024). Assistance may come in the form of the planned EU Data Union Strategy, which forms part of the AI Continent Action Plan (European Commission, 2025a,) which should hopefully make it easier to share and use data to the benefit of firms inside and outside AI regulatory sandboxes.

### *9.3 Appropriate Coordination between AI Regulatory Sandboxes*

As a network of national AI sandbox is established, effective mechanisms for coordination and sharing regulatory learning will prove critical to avoiding sandbox interpretative fragmentation and arbitrage impeding the effective achievement of the aims of the AI Act. This will require the AI Board to be dynamic in meeting the expectation of it to "facilitate cooperation and information-sharing among AI regulatory sandboxes" Art.66(k) AI Act). National competent authorities should be interacting with each other and also sharing their experiences through the medium of the AI Board. The AI Board's work will be augmented by the stakeholder advisory forum which can provide guidance to Member States that will assist with harmonisation of administrative practices around the functioning of AI regulatory sandboxes (Article 67(d) AI Act). Making sandbox exit reports centrally available on the EU platform (Art. 57(12) Ai Act) could also aid Member States' regulatory learning and future coordination.

### *9.4 Providing Clarity for Innovators Around Competing Supports*

Multiple supports exist within the ecosystem that are potentially welfare enhancing for AI Act compliance and thus compete for innovator attention. AI Innovators need to be able to make informed choices about which AI Act ecosystem supports would be of most benefit. It is both difficult and time-consuming to assess an expanding plethora of sometimes overlapping innovation supports on the path to market to decide which, if any, to avail of. A lack of clear, joined-up thinking around the various innovation supports and their purposes is frustrating for innovators and their advisors. The provision of supports therefore needs to be done in a manner that is coherent and comprehensible with clear and transparent functions and pathways and avoidance of unnecessary overlaps.

Navigation of available supports is not aided by the fact that the ecosystem is both shifting and multi-pronged. On testing and validation, how the AI regulatory sandbox meshes and connects with other innovation supports such as living labs, testbeds and hubs is not addressed in the AI Act (Pošćić & Martinović, 2022). As other testing and experimentation facilities for AI continue to come on stream to support the ecosystem such as EU-funded Sectorial AI Testing and Experimentation Facilities ("TEFs") (European Commission, 2025b) this point assumes increasing importance. The European Commission initially favoured the network of European Digital Innovation Hubs ("EDIHs") being used to provide compliance assistance to SMEs and start-ups(European Commission, 2020, p.23). However, the AI Continent Action Plan envisages the EDIHs as being more concerned with the nuts and bolts of driving AI adoption across enterprise rather than with regulatory compliance (European Commission, 2025a, pp.14-15).

Consequently, an EU audit of available support pathways concerned with testing and validation and regulatory compliance for guiding AI innovators, particularly SMEs is needed to clarify both supranational and national offerings in support of the AI ecosystem. Following on from this, disseminating accessible information on the ecosystem of supports at EU and national level would equip innovators to weigh up available alternatives rather than to devote time to applying for the scarce resources of an AI regulatory sandbox if it is not the best fit for the innovator's individual requirements. This would also help to avoid national sandboxes being overwhelmed be demand simply by virtue of being the most visible support in the ecosystem.

## 10. Conclusion

AI regulatory sandboxes have real potential to prove valuable in helping to ensure that AI systems being brought to market are robust and safe. Regulatory learning insights around

AI systems and AI Act compliance obtained in the sandbox environment could prove highly beneficial for both competent authorities and innovators admitted to AI regulatory sandboxes. However, the AI Act's vision for regulatory sandboxes is thinly sketched with a large measure of discretion is left to Member States. This means that the individual and collective success of AI regulatory sandboxes will rest on carefully curated design, implementation and coordination. Furthermore, EU and Member State sandbox design choices around AI regulatory sandboxes matter to their attractiveness to innovators versus substitutable ecosystem supports that assist with compliance en route to market comprising both other national AI regulatory sandboxes and other ecosystem supports.

As this early stage assessment of AI regulatory sandboxes has highlighted, the majority of Member States have not made significant headway around regulatory sandbox implementation. Presented with the triple challenge of capacity, coordination and attractiveness, attention is now needed from the European Commission, the AI Board and Member State competent authorities to hone the AI regulatory sandbox proposition to make it functional and cohesive. Ideally there should be a cohesive EU network of well-resourced and well-organised AI regulatory sandboxes providing a dynamic and supportive environment with clear operational parameters. This matters to outcomes. Arbitrage is likely to occur within the network of AI regulatory sandboxes under the AI Act, based on the perceived market utility or attractiveness of AI regulatory sandboxes to prospective providers of high-risk systems to the EU market as compared with other AI regulatory sandboxes and other supports and direct compliance routes. For the AI Act to achieve its aims around a harmonised rulebook, the potential for inconsistent regulatory guidance emanating from individual regulatory sandboxes that would cause national fragmentation, confusion and regulatory arbitrage jeopardising the EU's single market objectives must be guarded against. To address this challenge, the European Commission, the AI Board, the AI Office and national competent authorities, need to work responsively to ensure cohesion.

**Acknowledgments**

The author acknowledges with gratitude the feedback of the participants of the Milan Information Society Law Center's international conference on 11-12 June 2024.